\documentclass[11pt]{article}
\usepackage{moriond,epsfig}
\def\gsim{\mathrel{
\rlap{\raise 0.511ex \hbox{$>$}}{\lower 0.511ex
\hbox{$\sim$}}}}
\def\lsim{\mathrel{
\rlap{\raise 0.511ex \hbox{$<$}}{\lower 0.511ex
\hbox{$\sim$}}}}

\def\be{\begin{equation}}
\def\ee{\end{equation}}
\def\bea{\begin{eqnarray}}
\def\eea{\end{eqnarray}}

\begin{document}
\vspace*{4cm}
\title{SCIENTIFIC POTENTIAL OF EINSTEIN TELESCOPE}

\def\addon{\vskip0.1cm}
\author{
B\,Sathyaprakash$^{18}$,
M\,Abernathy$^{3}$,
F\,Acernese$^{4,5}$,
P\,Amaro-Seoane$^{33,46}$,
N\,Andersson$^{7}$,
K\,Arun$^{8}$,
F\,Barone$^{4,5}$,
B\,Barr$^{3}$,
M\,Barsuglia$^{9}$,
M\,Beker$^{45}$
% ...
}
\address{$^{1}$\,INFN, Sezione di Perugia, I-6123 Perugia, Italy  
\addon{$^{2}$\,European Gravitational Observatory (EGO), I-56021 Cascina (Pi), Italy}
\addon{$^{3}$\,SUPA, School of Physics and Astronomy, The University of Glasgow, Glasgow, G12\,8QQ, UK}
\addon{$^{4}$\,INFN, Sezione di Napoli, Italy}
\addon{$^{5}$\,Universit\`{a} di Salerno, Fisciano, I-84084 Salerno, Italy}
\addon{$^{6}$\,Max--Planck--Institut f\"{u}r Gravitationsphysik and Leibniz Universit\"{a}t Hannover, D-30167 Hannover, Germany}
\addon{$^{7}$\,University of Southampton, Southampton SO17\,1BJ, UK}
\addon{$^{8}$\,Chennai Mathematical Institute, Siruseri 603103 India}
\addon{$^{9}$\,AstroParticule et Cosmologie (APC), CNRS; Observatoire de Paris, Universit\'{e} Denis Diderot, Paris VII, France}
\addon{$^{10}$\,VU University Amsterdam, De Boelelaan 1081, 1081 HV, Amsterdam, The Netherlands}
\addon{$^{11}$\,Universit\'{e} Nice \textquoteleft Sophia--Antipolis\textquoteright, CNRS, Observatoire de la C\^ote d'Azur, F-06304 Nice, France}
\addon{$^{12}$\,Washington State University, Pullman, WA 99164, USA}
\addon{$^{13}$\,INFN, Sezione di Pisa, Italy}
\addon{$^{14}$\,Astronomical Observatory, University of warsaw, Al Ujazdowskie 4, 00-478 Warsaw, Poland}
\addon{$^{15}$\,Universit\`{a} di Napoli \textquoteleft Federico II\textquoteright, Complesso Universitario di Monte S. Angelo, I-80126 Napoli, Italy}
\addon{$^{16}$\,University of Birmingham, Birmingham, B15 2TT, UK}
\addon{$^{17}$\,INFN, Sezione di Genova, I-16146 Genova, Italy}
\addon{$^{18}$\,Cardiff University, Cardiff, CF24 3AA, UK}
\addon{$^{19}$\,INFN, Sezione di Roma Tor Vergata I-00133 Roma, Italy}
\addon{$^{20}$\,Universit\`{a} di Roma Tor Vergata, I-00133, Roma, Italy}
\addon{$^{21}$\,Moscow State University, Moscow, 119992, Russia}
\addon{$^{22}$\,INFN, Laboratori Nazionali del Gran Sasso, Assergi l'Aquila, Italy}
\addon{$^{23}$\,LIGO, California Institute of Technology, Pasadena, CA 91125, USA}
\addon{$^{24}$\,INFN, Gruppo Collegato di Trento, Sezione di Padova; Istituto di Fotonica e Nanotecnologie, CNR-Fondazione Bruno Kessler, I-38123 Povo, Trento, Italy}
\addon{$^{25}$\,Laboratoire des Mat\'{e}riaux Avanc\'{e}s (LMA), IN2P3/CNRS, F-69622 Villeurbanne, Lyon, France}
\addon{$^{26}$\,Institute of Astronomy, University of Cambridge, Madingley Road, Cambridge, CB3 0HA, UK}
\addon{$^{27}$\,Theoretical Astrophysics (TAT) Eberhard-Karls-Universit\"at T\"ubingen, Auf der Morgenstelle 10, D-72076 T\"{u}bingen, Germany}
\addon{$^{28}$\,INFN, Sezione di Firenze, I-50019 Sesto Fiorentino, Italy}
\addon{$^{29}$\,Universit\`{a} degli Studi di Urbino \textquoteleft Carlo Bo\textquoteright, I-61029 Urbino, Italy}
\addon{$^{30}$\,INFN, sezione di Padova, via Marzolo 8, 35131 Padova, Italy }
\addon{$^{31}$\,University of Minnesota, Minneapolis, MN 55455, USA}
\addon{$^{32}$\,Friedrich--Schiller--Universit\"{a}t Jena PF, D-07737 Jena, Germany}
\addon{$^{33}$\,Max Planck Institute for Gravitational Physics (Albert Einstein Institute) Am M\"{u}hlenberg 1, D-14476 Potsdam, Germany}
\addon{$^{34}$\,INFN, Sezione di Roma 1, I-00185 Roma, Italy}
\addon{$^{35}$\,
NSF Astronomy and Astrophysics Postdoctoral Fellow, MIT Kavli 
Institute, Cambridge, MA 02139}
\addon{$^{36}$\,University of Mississippi, University,  MS 38677, USA}
\addon{$^{37}$\,LAPP-IN2P3/CNRS, Universit\'{e} de Savoie, F-74941 Annecy-le-Vieux, France}
\addon{$^{38}$\,The Pennsylvania State University, University Park, PA 16802, USA}
\addon{$^{39}$\,Caltech--CaRT, Pasadena, CA 91125, USA}
\addon{$^{40}$\,Universit\`{a} di Pisa, I-56127 Pisa, Italy}
\addon{$^{41}$\,Universit\`{a} \textquoteleft La Sapienza\textquoteright, I-00185 Roma, Italy}
\addon{$^{42}$\,INFN, Sezione di Roma Tre and Universit\`{a} di Roma Tre, Dipartimento di Fisica,
I-00146 Roma, Italy}
\addon{$^{43}$\,Universit\`{a} degli Studi di Firenze, I-50121, Firenze, Italy}
\addon{$^{44}$ Departament de Fisica, Universitat de les Illes Balears,
Cra. Valldemossa Km. 7.5, E-07122 Palma de Mallorca, Spain}
\addon{$^{45}$ Nikhef, Science Park 105, 1098 XG Amsterdam, The Netherlands}
\addon{$^{46}$ Institut de Ci{\`e}ncies de l'Espai (CSIC-IEEC), Campus UAB, Torre C-5, parells, $2^{\rm na}$ planta, ES-08193, Bellaterra, Barcelona, Spain}
\addon{$^{46}$ KFKI Research Institute for Particle and Nuclear Physics, Budapest, Hungary}
}



\maketitle\abstract{
Einstein gravitational-wave Telescope (ET) is a design study funded by
the European Commission to explore the technological challenges of and
scientific benefits from building a third generation gravitational wave
detector. The three-year study, which concluded earlier this year,
has formulated the conceptual design of an observatory that can support
the implementation of new technology for the next two to three decades.
The goal of this talk is to introduce the audience to the overall aims
and objectives of the project and to enumerate ET's potential to influence
our understanding of fundamental physics, astrophysics and cosmology.}

\section{Introduction}
\label{sec:introduction}
Interferometric gravitational wave (GW) detectors, Laser Interferometer
Gravitational-Wave Observatory (LIGO) in the US, 
Virgo, 
GEO600 
and TAMA, 
have successfully operated at design 
sensitivities for a year or more~\cite{Abbott2009,Accadia2010}. They have 
demonstrated that it is possible to build and run these highly sensitive 
instruments with a large duty cycle~\cite{GEO600Grote2008}. While no 
signal has so far been observed in any of these detectors, their data 
have been used to break new ground on several astronomical 
sources~\cite{Abbott:2008fx,Abbott:2009ws,LSC:GRB070201,LSC:S5GRBBurst}.

The network of advanced detectors, which includes 
advanced LIGO~\cite{AdvancedLIGOReference2009}, advanced Virgo~\cite{AdV2009}, 
Large Cryogenic Gravitational Telescope~\cite{LCGTProject} (to be built in
Kamioka mines in Japan) and GEO-HF~\cite{Willke2006} (GEO High Frequency),
is expected to make the first direct detection of GW sometime during this
decade. This will be a new milestone for observational astronomy that will 
facilitate the study of formations and interactions of neutron stars (NSs) and
black holes (BHs) in the Universe. 

Direct detection of GW will allow the study of phenomena associated with strong
gravitational fields and relativistic gravity that are otherwise not accessible
to us. They will allow new tests of general theory of relativity in
regimes where one might expect to see departure from standard predictions.
The study of GW sources will by itself establish as a new field of observational
astronomy. However, there is much more to be benefitted beyond the mere study
of phenomena associated with GW sources. Just as stars, GW sources are
markers in space, sometimes with precisely known distances. They could,
therefore, serve to study the structure and dynamics of the Universe and hence
a new tool for cosmology. 

Advanced detectors will study NSNS, NSBH and BHBH binaries
at distances of 200 Mpc, 600 Mpc and 3 Gpc, respectively, within 
which the nominal event rates are about 40 per year for NSNS binaries
and similar, but much more uncertain, rates for NSBH and BHBH 
binaries~\cite{2010CQGra..27q3001A}. The signal-to-noise ratio (SNR) for most of 
the sources detected by advanced detectors will be around 10. This should already 
make it possible to carry out a number of accurate measurements that will 
impact fundamental physics and astrophysics~\cite{lrr-2009-2}. 
For instance, it should be possible to measure the Hubble constant to within 1\%
if NSNS and NSBH binaries are progenitors of short-hard gamma ray 
bursts (GRBs) and confirm the presence of tails of gravitational 
waves by observing BHBH mergers~\cite{Blanchet:1994ez}. 

Third generation detectors, such as the Einstein Telescope (ET), will have 
ten times greater SNR for the same events and their reach will increase
to $z\simeq 2,$ for NSNS binaries, $z\simeq 6$ for NSBH binaries 
and $z\simeq 17$ for BHBH binaries (cf.\, Fig.\,\ref{fig:reach}). 
They will help address a variety of issues associated with phenomena 
that have remained as enigmas for several years to decades after 
their initial discovery. More than anything else, ET might well unveil 
new physics beyond the standard models of particle physics and cosmology.

The purpose of this talk is to discuss the science potential
of ET and how it will be a powerful new tool for observing phenomena 
associated with strong field, relativistic gravity.  The design study 
has already provided useful insight on what really will be the benefit 
of building a third generation GW detector~\cite{DSD}. However, 
the {\em full} science potential of ET and the challenges posed 
by science exploitation, remain unexplored. Yet what has been 
investigated is already very exciting and should
provide the impetus for further studies. The talk will begin with a 
brief description of the technical aspects of the design
and different sensitivity options, followed by a discussion of ET's 
science potential.

\section{ET Sensitivity}
\label{sec:sensitivity}

The ET design study was commissioned by the European Commission to scope
out the technological feasibility of building a 3rd generation detector
and to explore its science potential. The study team set itself the goal
of designing a detector that is better than advanced detectors ten times
in strain sensitivity and reaches down to 1 Hz rather than the 10-20 Hz
low frequency limit of advanced detectors.  It was soon realized that 
the infrastructure, in which advanced detectors will have been housed 
for more than 20 years since their inception, will be highly inadequate 
in realizing the sensitivity of a 3rd generation detector. ET will be
more than just a detector; it will be a facility that will house 
a 3rd generation observatory but with infrastructure that can support
new designs and improvements for several decades.

A factor ten in strain sensitivity is achieved by a combination of
increased arm lengths (10 km arms as opposed to 3-4 km arms afforded 
by the current infrastructures), seismically quieter underground 
environments to mitigate seismic noise, higher arm cavity laser powers 
to confront photon shot noise and cryogenic mirrors cooled down to 10 K
to reduce thermal noise.

\subsection{Arm lengths and topology}
\begin{figure*}
\centering
\includegraphics[angle=0,width=0.45\textwidth]{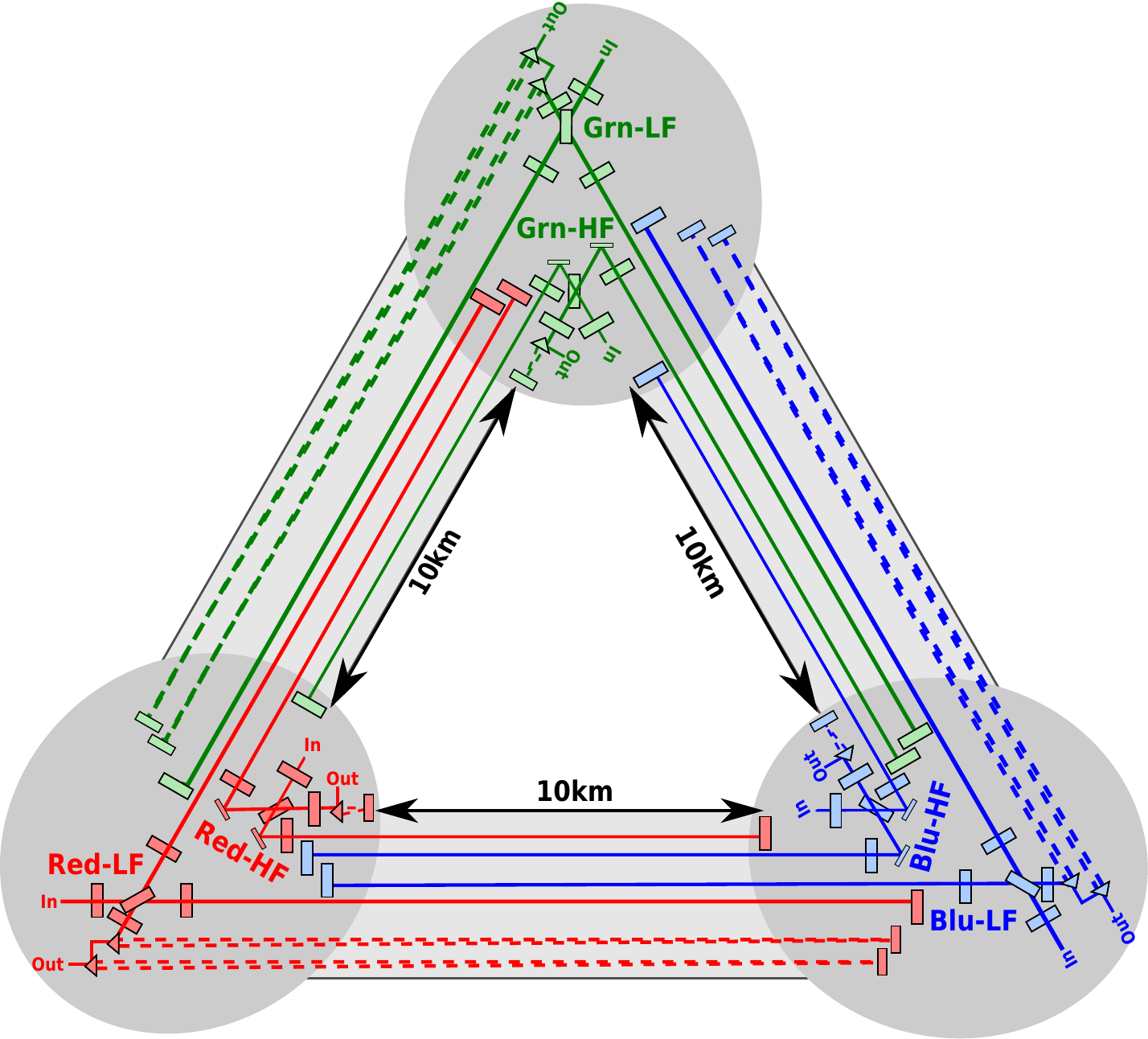}
\hskip0.1\textwidth
\includegraphics[angle=0,width=0.28\textwidth]{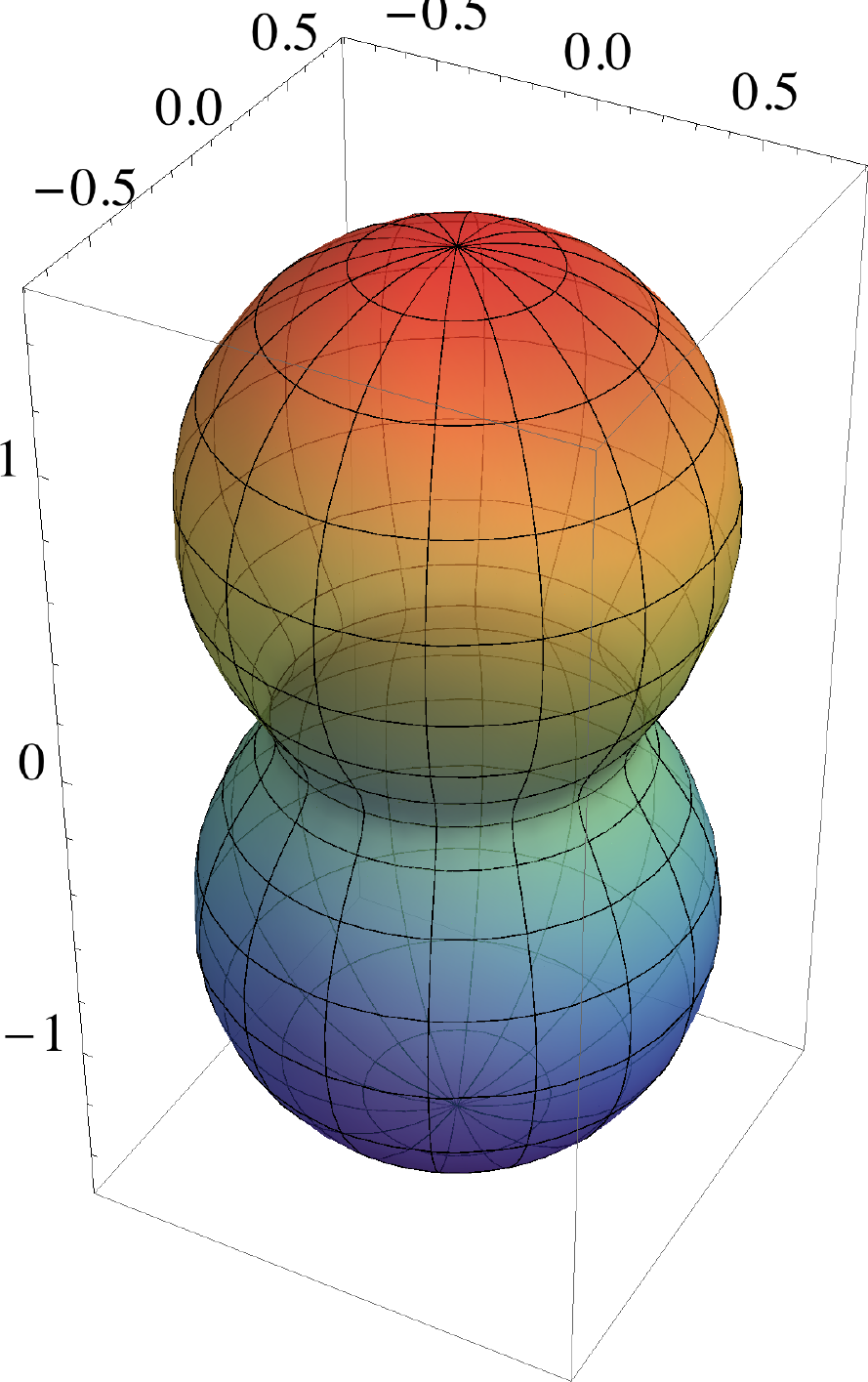}
\caption{
{\em Left}: 
Schematic full view of the optical layout of the ET Observatory. It consists 
of 3 pairs of km-scale interferometers positioned such that they form 
a triangular shape. Each interferometer pair represents one wide-band 
detector, in which one interferometer is optimized for gravitational 
waves at low frequencies (i.e., $<100\,\rm Hz$) and the other for high 
frequencies (i.e., $>100\,\rm Hz).$ {\em Right:} The joint antenna 
pattern of the three interferometers to sources from around the sky. 
ET has virtually full sky coverage.
\label{fig:et}}
\end{figure*}

In the long-wavelength approximation, the strain sensitivity of an 
interferometer increases in direct proportion to the length of its arms. 
The arm lengths of current (large) detectors is either 3 or 4 km. 
The strain sensitivity of an interferometer with 10 km arms will be 
2.5 to 3 greater. Current ground-based detectors are L-shaped interferometers
since an opening angle of 90 degrees maximizes their sensitivity. However,
careful considerations taking into account continuous operation, 
ability to resolve the two independent wave polarizations and minimizing 
the infrastructure costs, favours the construction of a triangular 
configuration. 

The advantage of a triangular topology is that each side of the
triangle can be deployed twice to build, in effect, three 
V-shaped interferometers with an opening angle of 60 degrees 
and rotated relative to each other by 120 degrees (see the panel 
on the left in Fig.\,\ref{fig:et}). An opening angle of 60 degrees 
means that the sensitivity reduces to $\sqrt{3}/2$ that of an L-shaped 
detector; the three detectors in the triangle enhance the sensitivity
by a factor of $\sqrt{3}$ and so an overall gain in sensitivity of 3/2.
The panel on the right in Fig.\,\ref{fig:et} shows the antenna pattern
of the triangular network. The triangular ET has virtually complete 
sky coverage and it has no blind spots. Its reach to sources lying
in the plane of the triangle will be a third of its reach to sources 
lying overhead!

The three V-shaped interferometers are, 
of course, equivalent in sensitivity to two L-shaped interferometers
with arms that are only three-quarters in size of the triangular arms
and rotated relative to each other by 45 degrees. 
However, the responses of the three 
detectors in a triangle can be used to construct a {\em null stream} 
that is {\em not} possible with the two L-shaped interferometers. It 
turns out that the sum of the responses of the three detectors in
a triangle (for that matter any closed topology) is completely 
devoid of any gravitational wave. This is the closest that one can get to
measuring the ``dark current" in interferometers. The 
null stream will be an invaluable tool to characterize the background. 

\subsection{Going underground}
\begin{figure*}
\centering
\includegraphics[width=0.49\textwidth]{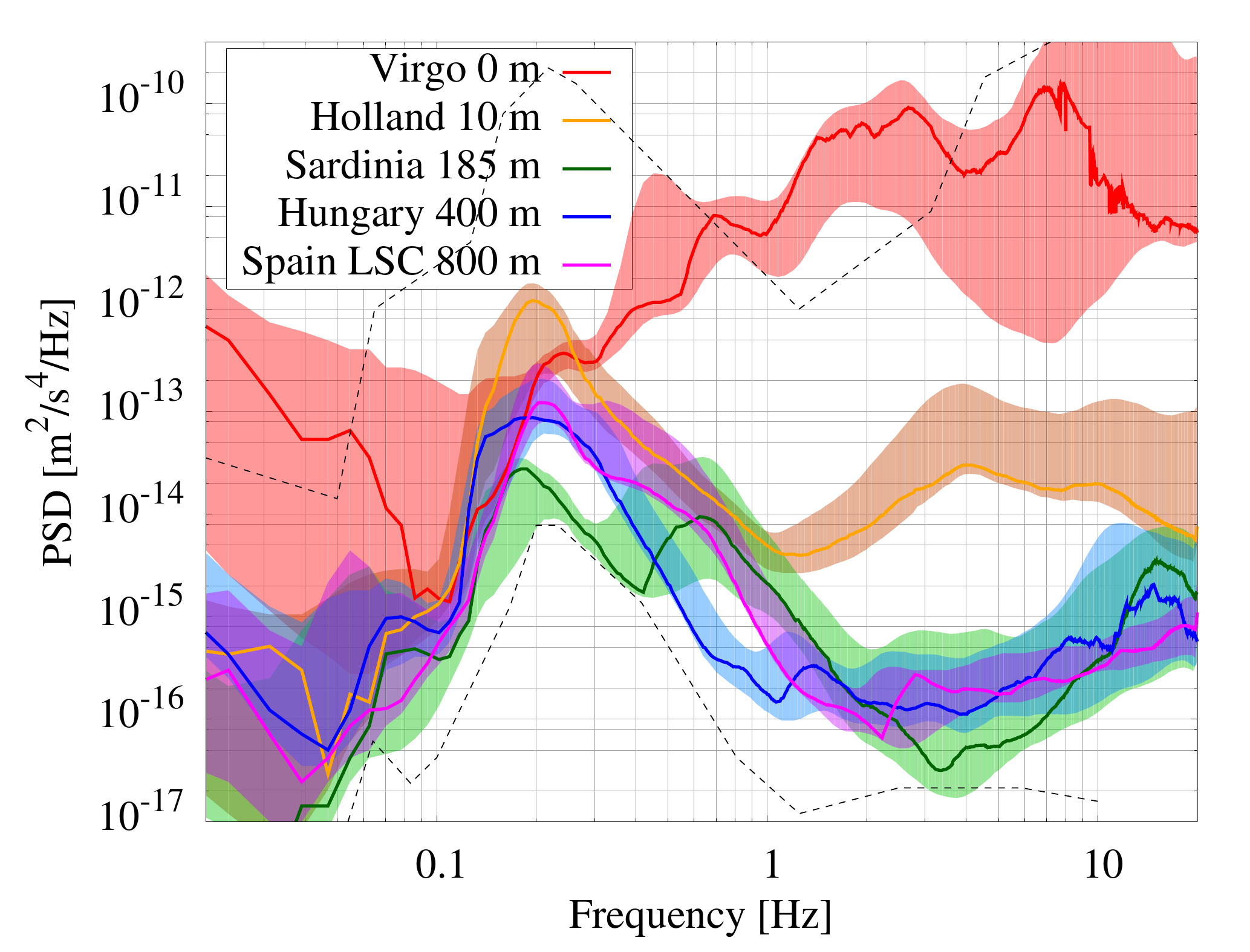}
\includegraphics[width=0.48\textwidth]{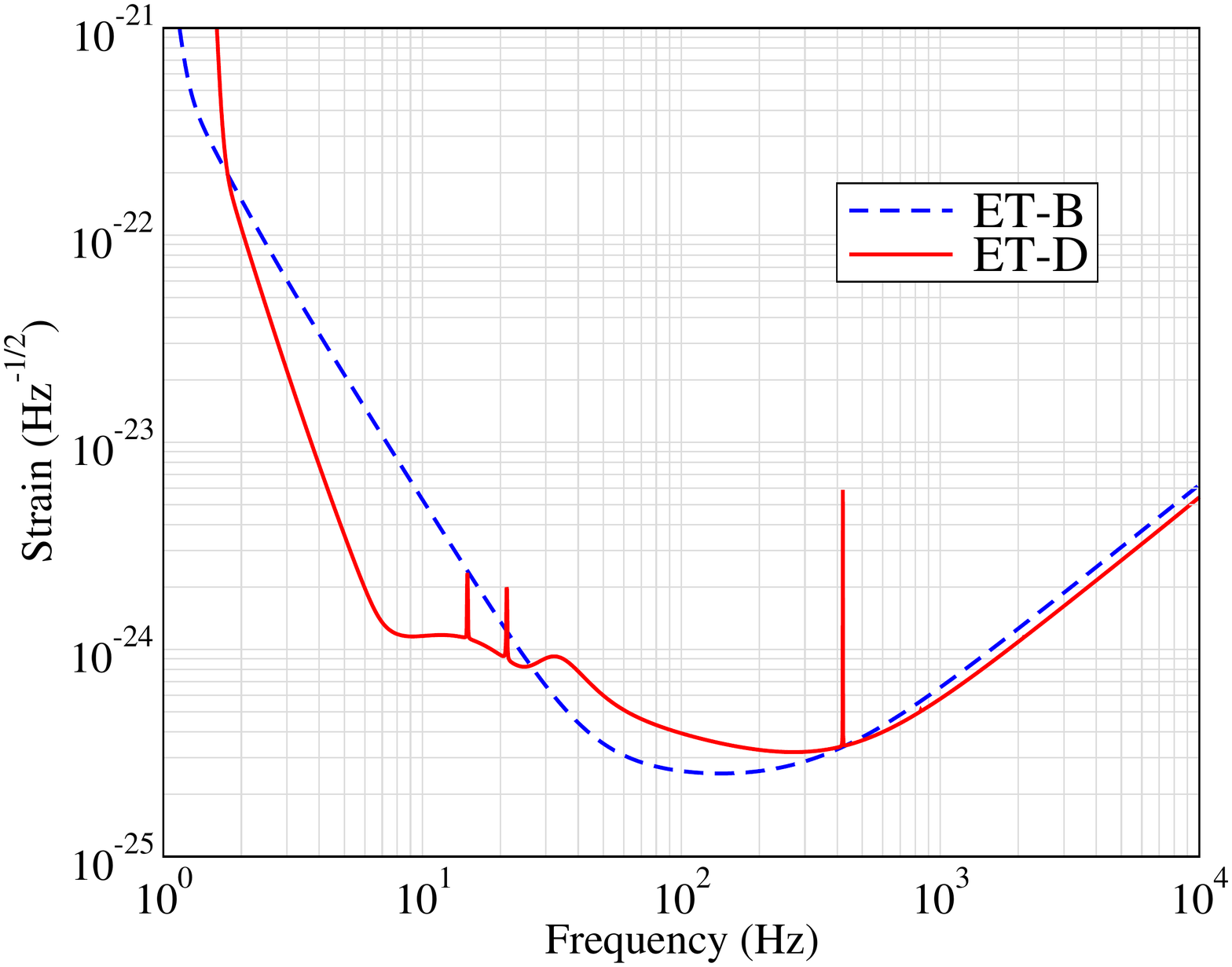}
\caption{
{\em Left:}
The spectrum of horizontal motion over one-week period at Cascina, 
where Virgo is located, is compared to those measured at several 
underground locations in Europe.
The solid lines correspond to the mode, while the upper and lower 
limits of the transparent regions are the PSD levels that weren't 
exceeded for 90\% and 10\% of the time respectively.
{\em Right:} The sensitivity of ET for the xylophone 
configuration, ET-D, is compared with that of a conventional 
configuration that achieves broadband sensitivity with a single 
interferometer, ET-B.
\label{fig:seis}}
\end{figure*}

Achieving good low frequency sensitivity requires mitigation of gravity
gradients that are far too high on ground. They can be circumvented
either by getting into space (the option pursued by the Laser 
Interferometer Space Antenna) or by going underground. To be useful,
any underground site must be seismically quiet. Figure \ref{fig:seis}
shows the seismic noise in several European underground sites compared
to the seismic noise at Cascina, where Virgo is located. Clearly,
underground environments could be several orders of magnitude quieter than
ground-based ones. 

Achieving a good sensitivity over a broad frequency range from 1 Hz to 10 kHz
with the same technology is impractical.
The technology required for better high frequency (i.e. $> 100\,\rm Hz$) 
sensitivity -- higher laser powers -- is in direct conflict with that 
required for improving the low frequency (i.e. $< 100\,\rm Hz$) sensitivity,
namely low thermal and radiation pressure noises. Thus it is not 
prudent to build a single detector that meets the design goal in the
entire frequency band. Instead, the design study concluded that it
is best to build separate interferometers for the low and high
frequency regions. 

\subsection{Megawatt lasers, squeezed light and cryogenic mirrors}

The key to high frequency sensitivity is high laser power. Above $\sim
100\, \rm Hz,$ the main source of noise is the photon shot noise, which
can be reduced by simply using as high a power in the cavity as possible.
ET aims to achieve the required 3 MW of power by using inherently
more powerful input lasers (500 W as opposed to the 180 W in advanced 
interferometers). Furthermore, the use of non-classical light, squeezed 
light, leads to further improvement in sensitivity~\cite{McKenzie2004}.  
Indeed, ET design assumes a squeezing factor of 10 dB, which is equivalent 
to shot noise reduction resulting from an increase in laser power of 
a factor of 10.

Although, higher laser power works well at frequencies above 100 Hz,
it has the adverse effect of worsening the sensitivity in the 10-100 Hz.
This is due to enhanced thermal noise in mirror substrates and coating. 
Thus, it is not sensible to achieve the sensitivity goal over the entire
band with a single interferometer. The current thinking is to build a pair
of interferometers in each V of the triangle, one using high laser 
powers and the other with lower laser powers and cryogenic mirrors to 
mitigate thermal noise.

Figure \ref{fig:seis}, right panel, plots the strain sensitivity 
(per $\sqrt{\rm Hz}$) for the xylophone configuration 
ET-D~\cite{Hild:2010id}~\footnote{The data for the sensitivity curves 
can be found at {\tt http://www.et-gw.eu/etsensitivities.}}.
The xylophone configuration deploys a pair of interferometers 
to achieve good broadband sensitivity. Also shown is the sensitivity
of a conventional configuration ET-B~\cite{HildETconventional},
that has only one interferometer in each V of the triangle. 
Apart from the frequency range from 20 to 200 Hz where ET-B 
is slightly better than ET-D, the xylophone configuration quite
significantly wins over ET-B in the low frequency 
range.

\section{ET's science objectives}
\begin{figure*}
\centering
\includegraphics[width=0.48\textwidth]{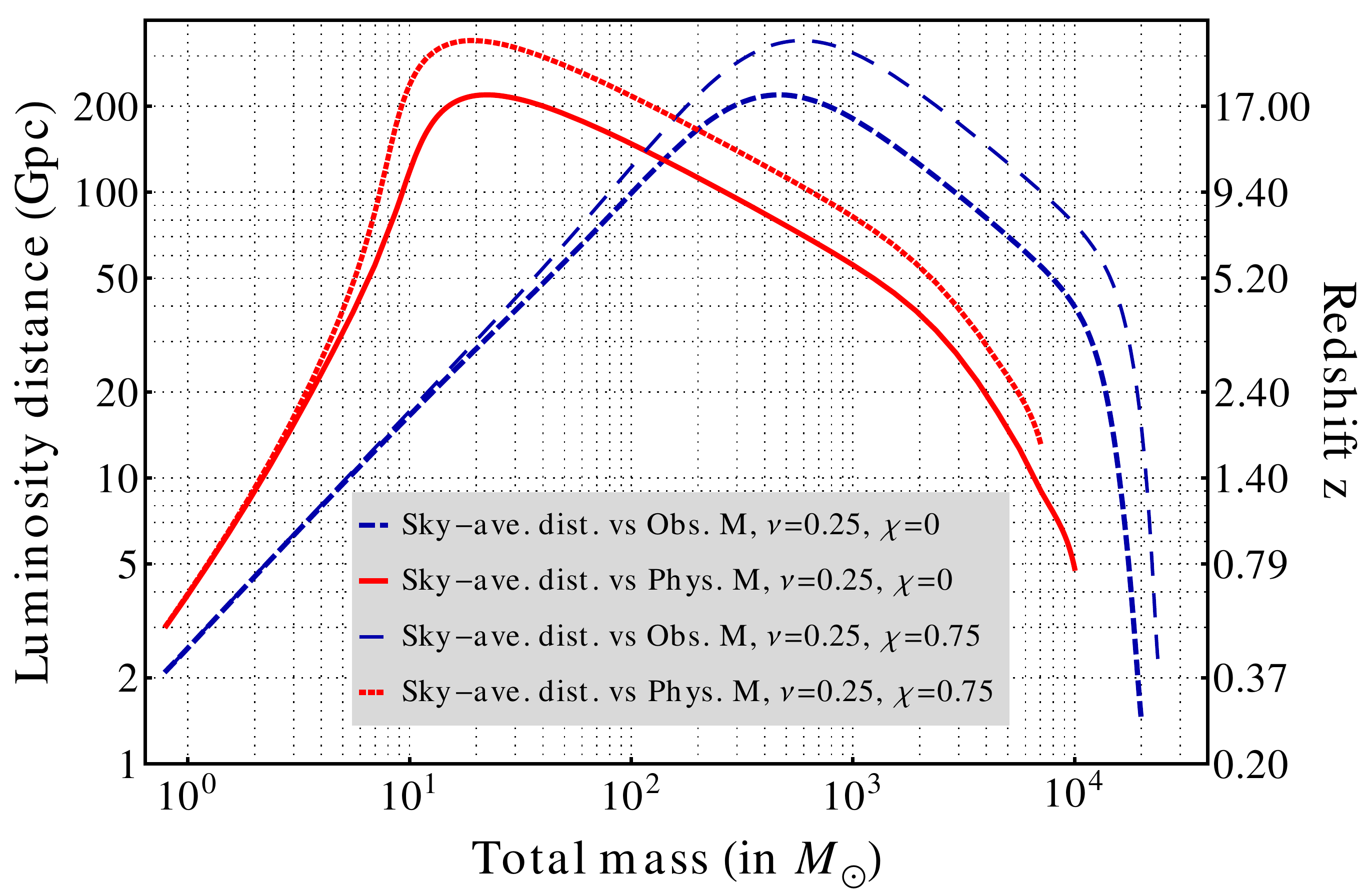}
\includegraphics[width=0.48\textwidth]{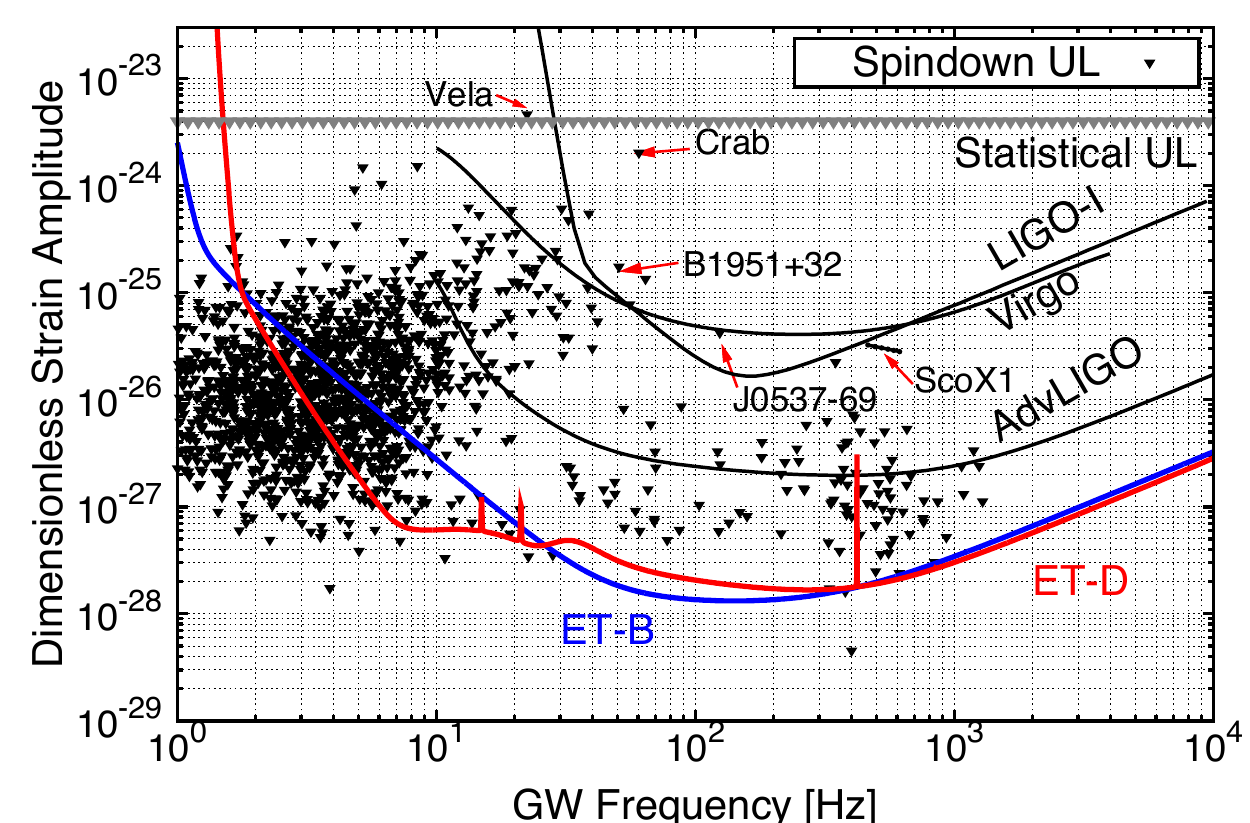}
\caption{
Plots show the distance reach of ET for compact binary mergers as 
a function of the total mass (left) and its sensitivity to GWs from 
known pulsars (right). See the text for details. 
\label{fig:reach}}
\end{figure*}

ET's distance reach for inspiralling and merging black holes 
for ET-B sensitivity is shown in the left panel of Fig.\,\ref{fig:reach}.
The long- and short-dashed curves correspond to the observed total 
mass $M_{\rm obs}$ and the solid and dotted curves correspond to 
the intrinsic total mass $M_{\rm int};$ the two are related by 
$M_{\rm int}=M_{\rm obs}/(1+z).$ The solid and short-dashed 
curves are for non-spinning binaries consisting of two equal masses, 
while the dotted and long-dashed curves are the same except that 
the component black holes are both assumed to have a dimensionless spin 
magnitude of 0.75. 

It is immediately apparent that ET will be sensitivity
to BHBH binaries of intrinsic total mass 10-20$M_\odot$ at a redshift 
of $z\sim 10$ and beyond. NSNS binaries could be seen when the star formation
in the Universe is at its peak at $z\sim 2.$ NSBH binaries comprising of
a 1.4 $M_\odot$ NS and a 10 $M_\odot$ BH can be detected from redshifts of 
at least $z\sim 6.5.$ Together with the fact that the inspiral phase of
compact binaries are standard sirens~\cite{Schutz86} means that
ET will be able to explore not only the properties of the sources themselves
but can also act as a tool to probe the properties of the Universe.
Intermediate mass black holes of intrinsic total mass in the range 
$10^2$-$10^4 M_\odot$ can be seen in the redshift range of 1 to 10,
thus offering a unique probe to uncover a host of questions related to
their existence and their role in the formation and evolution of galaxies.

Also shown in Fig.\,\ref{fig:reach}, right panel, are the sensitivities of 
initial LIGO, Virgo, advanced LIGO and ET (two versions, ET-B and ET-D), to 
continuous waves 
from rotating, asymmetric neutron stars, for an integration period 
of five years.  Inverted black triangles give the upper limit on 
the amplitude of GW of known pulsars derived by assuming that their
observed spin-down rate is entirely due to the emission of GW -- Vela,
Crab, B1951+32 and J0537-69 being specific examples. The horizontal
line shows the limit on the amplitude of GW from pulsars obtained
from statistical arguments.  ET-D (red curve) will be sensitive 
to intrinsic GW amplitudes greater than $h\sim 10^{-27}$ in the 
frequency range 6 Hz to 3 kHz, and a factor 3 better in the range 
20 Hz to 1 kHz. It is particularly important that ET is able to
reach sensitivity levels that are two to four orders of magnitude
lower than the spin-down limits, where one might have a real chance
of detecting a signal.

The rest of this paper enumerates ET's science goals in fundamental
physics, astrophysics and cosmology.

\subsubsection{Probing fundamental physics with ET}
\subsection{Is the nature of gravitational radiation as predicted by
Einstein's theory? }
ET will allow a test of the wave generation formula beyond the quadrupole
approximation~\cite{TestGR:2010}.  It could accurately measure the GW propagation
speed by coincident observation of GW and EM radiation from NSNS binary
coalescences at $z\sim 2$ and constrain the graviton mass~\cite{AW09}.

\subsection{Are black hole spacetimes uniquely given by the Kerr geometry?}
By measuring different quasi-normal modes, ET will test if
the spacetime geometry of a BH is uniquely described by its mass and 
spin~\cite{Kamaretsos:2011}.
Additionally, ET can measure the multipole moments of a source from the
radiation emitted as a stellar-mass BH spirals into an
intermediate-mass BH and confirm if the different moments
depend only on the massive BH's mass and 
spin~\cite{Huerta:2011,AmaroSeoaneSantamaria:2009}.

\subsection{What is the physics of gravitational collapse?}
ET can study supernovae and explore if they leave behind a massive
object that is trapped inside an event horizon or lead to a naked
singularity, or some other exotic object. ET could well reveal a new
class of objects and phenomena, for instance 
\emph{silent supernovae}~\cite{Woosley:1992} and other 
gravitationally unstable transients.

\subsection{What is the equation of state of matter at
supra-nuclear densities as might be found in NS cores?}
The equation of state (EoS) of NSs affects the late-time evolution of
NSNS and NSBH binaries. By matching the observed radiation from
the coalescence of such sources to theoretical predictions ET will
deduce the EoS of NS cores~\cite{Read:2009bns,2011GReGr..43..409A}.

\subsection{What is the maximum mass of a neutron star?}
The maximum mass of a white dwarf is $\simeq 1.4\,M_\odot$ as determined
by the electron degeneracy pressure.  The maximum mass of a NS is an
additional test of the nature of matter at extremely high densities;
it is currently unknown and should be determined by ET by accurately
constructing their mass function from millions of NSNS 
binaries~\cite{Sathyaprakash:2011a}.


\subsubsection{ET's impact on astrophysics and multimessenger astronomy}
\subsection{What is the mass function of BHs and NSs and
their redshift distribution?}
ET will measure masses and spins of millions of NSs and BHs
in binary systems and will thereby obtain a census of these 
objects as a function of redshift. This will be a very valuable 
tool for understanding a host of questions in astronomy related 
to redshift evolution of compact objects~\cite{VanDenBroeck:2006ar}.

\subsection{What are the progenitors of gamma-ray bursts?}
GRBs are the most luminous electromagnetic sources in the
Universe. While advanced detectors might provide some clues as to their
origin, ET will provide a large statistical sample of events that
could be used to understand GRB progenitors and to test their astrophysical 
models~\cite{Sathyaprakash:2011a}.

\subsection{How do compact binaries form and evolve?}
The process by which main sequence binary stars evolve into compact
binaries (that is, NSNS, NSBH and BHBH)
could be understood by ET's observation of millions of coalescing
binaries with different masses, mass ratios and spins and mapping the
observed population to astrophysical models~\cite{StarTrack}.

\subsection{What is the physical mechanism behind supernovae and how asymmetric
is the gravitational collapse that ensues?}
Supernovae are complex processes whose modelling requires many
different inputs, including relativistic magneto-hydrodynamics, 
general relativity and nuclear and particle physics~\cite{ott:09b}.
ET's observation of supernovae in coincidence with the detection of
neutrinos could provide the data necessary to constrain models and help
understand the process by which stars collapse to form NSs and BHs.

\subsection{Do relativistic instabilities occur in young NSs and if so
what is their role in the evolution of NSs?}
Non-linearities of general relativity could cause instabilities in
NSs that lead to parametric amplification of GWs.
ET's observations of the formation of NSs can explore if such
instabilities occur in young NSs and how that might affect
their spin frequencies~\cite{2011GReGr..43..409A}.

\subsection{Why are spin frequencies of NSs in low-mass X-ray binaries bounded?}
ET will verify if gravitational radiation back-reaction torque is responsible
for the observed upper limit on NS spin frequencies in low-mass 
X-ray binaries~\cite{Bildsten:1998ey}.

\subsection{What is the nature of the NS crust and its interaction
with the core?}
ET should detect NS ellipticities that are $\rm few \times 10^{-10}$
(for sources within a distance of 1 kpc) or larger depending on their 
spin frequency and their distance from earth.  Such observations 
can be used to deduce the property of NS crusts. ET might also detect GWs
that are expected to be emitted when pulsars glitch and magnetars flare and thereby
help understand crust-core interaction that is believed to transfer angular
momentum from the core to crust~\cite{Ruderman:1976}.

\subsection{What is the population of GW sources at high redshifts?}
A large population of point sources would produce a confusion background that would
be detectable by ET if the energy density of the background is large enough.
Detection of confusion backgrounds can be used to understand the nature and
population of GW sources in the Universe.

\subsubsection{ET as a new cosmological tool}

\subsection{What are the luminosity distances of cosmological sources?}
Compact binaries are an astronomer's ideal {\em standard candles}
or, more appropriately, {\em sirens}.
Gravi\-tational wave observations can alone determine both the
apparent and absolute luminosity of a source and hence deduce their luminosity
distance. With ET, these self-calibrating
standard sirens can be used to calibrate the cosmic distance 
ladder~\cite{Sathyaprakash2009}.



\subsection{What is the EoS of dark energy and how does it vary with redshift?}
ET could observe thousands of coalescing NSNS and NSBH
systems in coincidence with optical or gamma-ray observations
and hence measure both the luminosity distance and
redshift. ET will, therefore, facilitate precision
measurement of the dark energy EoS and its variation with redshift~\cite{Zhao:2010sz}.

\subsection{How did the black holes at galactic nuclei form and evolve?}
ET can verify if seeds of galaxy formation were intermediate BHs
of hundreds to thousands of solar masses and map their merger history up to
redshifts of $z\sim 5$--15 depending on the total mass and mass ratio of
progenitor binaries~\cite{sgmv09a,AmaroSeoaneSantamaria:2009,Gair:2010dx}.

\subsection{What were the physical conditions in the primeval Universe and what
phase transitions occurred in its early history?}
Stochastic GW backgrounds could be produced by quantum
processes in the primordial Universe or during phase transitions in its
early history. ET will be sensitive to background densities $\rho_{\rm GW}
\sim 10^{-12}\,\rho_c,$ where $\rho_c$ is the critical density of the 
Universe~\cite{ET-MDC}.

\section{Conclusions}
This decade will see the construction and operation of second generation
interferometric detectors, pulsar timing arrays and results from the
Planck space mission. There is little doubt that we are at the verge of a
new era in astronomy that will witness the opening of the gravitational 
window for observing the Universe.  I hope this talk has convinced you
that the field promises to have a huge potential and that ET can 
not only help solve some of the enigmas in astronomy and cosmology but
push the frontiers of science into new avenues.

\section*{References}
\bibliography{ReferencesET}{}

\end{document}